\documentclass[10pt,conference]{IEEEtran}
\usepackage{amsmath,amsfonts}
\usepackage{graphicx,algorithm,algpseudocode}
\usepackage{cite}
\usepackage{xcolor}
\usepackage[font=small]{caption}
\usepackage{subcaption}
\usepackage{amsfonts,booktabs}
\usepackage{multirow,verbatim}
\usepackage{color, colortbl}
\usepackage{lipsum}
\usepackage{soul}
\usepackage[normalem]{ulem}
\usepackage{float}

\newcolumntype{L}[1]{>{\raggedright\let\newline\\\arraybackslash\hspace{0pt}}m{#1}}
\newcolumntype{C}[1]{>{\centering\let\newline\\\arraybackslash\hspace{0pt}}m{#1}}
\newcolumntype{R}[1]{>{\raggedleft\let\newline\\\arraybackslash\hspace{0pt}}m{#1}}

\IEEEoverridecommandlockouts
\allowdisplaybreaks

\begin{document}
\bstctlcite{IEEEexample:BSTcontrol} % this is to make sure that if there are same authors in consecutive references, they are not dashed out. 

\title{Targeted Attacks and Defenses for Distributed Federated Learning in Vehicular Networks}

\author{
\IEEEauthorblockN{Utku Demir\IEEEauthorrefmark{1}, Tugba Erpek\IEEEauthorrefmark{1},  Yalin E. Sagduyu\IEEEauthorrefmark{1}, Sastry Kompella\IEEEauthorrefmark{1}, Mengran Xue\IEEEauthorrefmark{2}} \\ \IEEEauthorrefmark{1}Nexcepta Inc., Gaithersburg, MD, USA \\
\IEEEauthorrefmark{2}RTX BBN Technologies, Cambridge, MA, USA
}

\maketitle
\thispagestyle{plain}
\pagestyle{plain}

\begin{abstract}
In emerging networked systems, mobile edge devices such as ground vehicles and unmanned aerial system (UAS) swarms  collectively aggregate vast amounts of data to make machine learning decisions such as threat detection in remote, dynamic, and infrastructure‐constrained environments where power and bandwidth are scarce. Federated learning (FL) addresses these constraints and privacy concerns by enabling nodes to share local model weights for deep neural networks instead of raw data, facilitating more reliable decision‐making than individual learning. However, conventional FL relies on a central server to coordinate model updates in each learning round, which imposes significant computational burdens on the central node and may not be feasible due to the connectivity constraints. By eliminating dependence on a central server, distributed federated learning (DFL) offers scalability, resilience to node failures, learning robustness, and more effective defense strategies. Despite these advantages, DFL remains vulnerable to increasingly advanced and stealthy cyberattacks. In this paper, we design sophisticated targeted training data poisoning and backdoor (Trojan) attacks, and characterize the emerging vulnerabilities in a vehicular network. We analyze how DFL provides resilience against such attacks compared to individual learning and present effective defense mechanisms to further strengthen DFL against the emerging cyber threats.

\end{abstract}


\section{Introduction}
\label{sec:introduction}
Networked systems increasingly depend on connected and autonomous agents such as in vehicular systems, driven by technologies like Dedicated Short-Range Communications (DSRC) and Cellular Vehicle-to-Everything (C-V2X)~\cite{9645261}. As vehicles rely more heavily on real-time data sharing and collaborative decision-making, they face growing vulnerabilities to cyber threats, system faults, and environmental disruptions~\cite{9447840}. Traditional centralized monitoring, hindered by latency, bandwidth limitations, and infrastructure dependence, struggles to meet the demands of timely anomaly detection in these dynamic settings. To ensure operational security and resilience, decentralized solutions are ultimately needed to detect and respond to threats at the edge, safeguarding mission-critical communications and maintaining system integrity.

%\textbf{Federated Learning:}
The security of vehicular networks, such as those linking autonomous ground vehicles, unmanned aerial systems (UASs), and command-and-control platforms, is critical for operational effectiveness and situational awareness. Federated learning (FL) offers a compelling approach to securing these networks by enabling decentralized model training across distributed nodes without exposing sensitive raw data~\cite{elbir2022federated}, thus aligning with strict confidentiality requirements inherent in various defense and commercial environments~\cite{10105168}. From the cybersecurity perspective, FL has been utilized for detecting misbehavior in safety messages~\cite{huang2024semi}, identifying anomalies within the Internet of Vehicles (IoV)~\cite{tham2023federated}, predicting vehicle trajectories under cyber threats~\cite{wang2023federated}, and detecting V2X misbehavior in 5G edge computing environments~\cite{yakan2023federated}.

%\textbf{Distributed Federated Learning:}
However, traditional FL often relies on a centralized coordinator to aggregate model updates, introducing potential single points of failure and communication bottlenecks that are unacceptable in contested or bandwidth-limited theaters of operation. Distributed federated learning (DFL) addresses these shortcomings by eliminating the need for a central aggregator and instead enabling peer-to-peer model sharing and collaborative consensus-building among nodes~\cite{shia022federated, demir2025distributedfederatedlearningvehicular}. This architecture enhances the resilience, scalability, and robustness of learning systems, allowing them to adapt rapidly to emerging threats such as cyber intrusions, jamming, or spoofing, while maintaining operational security~\cite{10795238}. % and mission continuity~\cite{10795238}.

\begin{figure}[t!]
\centering
  \includegraphics[width=0.75\linewidth]{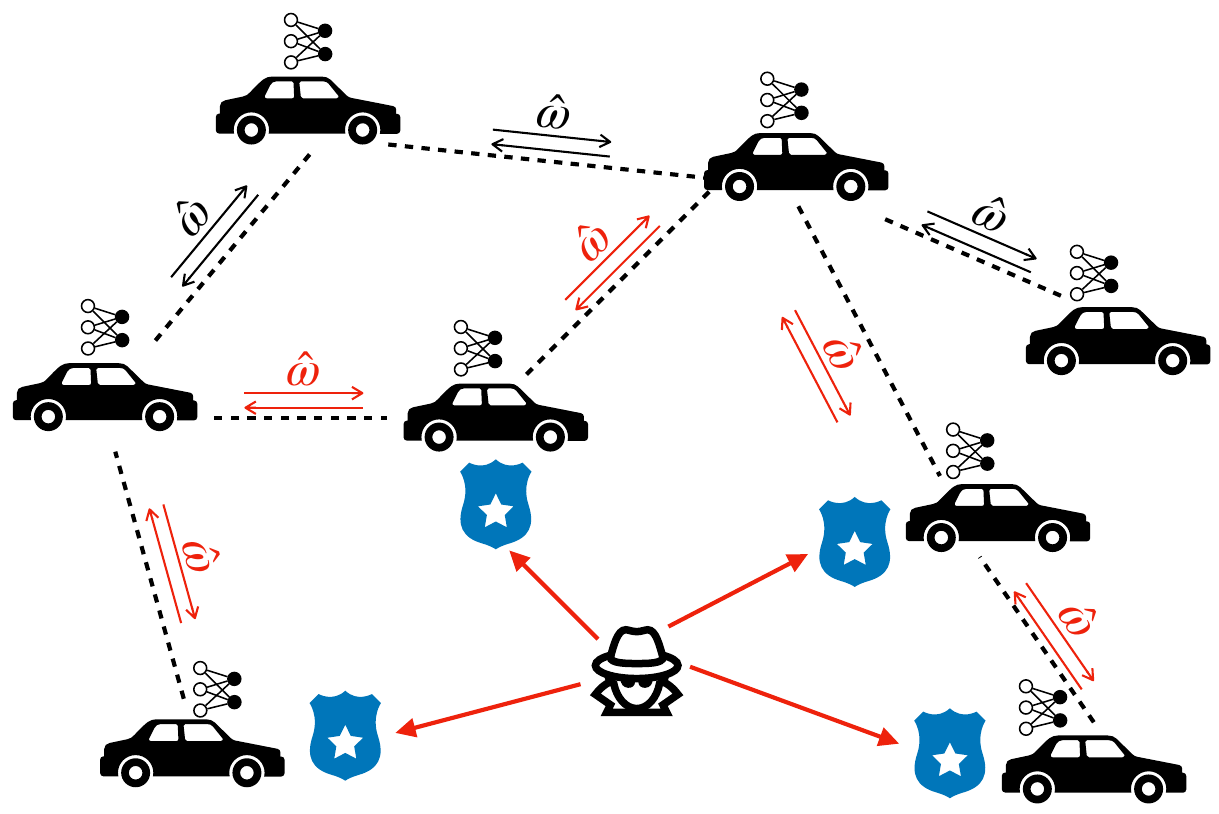}
    \caption {Poisoning attack on DFL, where false information is injected to the learning process.}
    %\vspace{-2mm}
  \label{fig:intro}
\end{figure}

While the decentralized structure of DFL reduces reliance on central infrastructure, it is still susceptible to multilayered security risks. As a wireless attack, jamming has been studied to disrupt communications for model exchange in both FL and DFL systems \cite{shi2022jamming,shi2023jamming, shi2022launch, demir2025distributedfederatedlearningvehicular}. From the machine learning security point of view, adversaries may launch various attacks on FL, in which compromised vehicles inject false information into the learning process by altering both the features of the training data and the corresponding labels~\cite{sagduyu2023securing, 10136777, demir2025distributedfederatedlearningvehicular}, as illustrated in Fig.~\ref{fig:intro}. Data poisoning might corrupt local datasets, spreading incorrect information and ultimately diminishing the global model’s performance. A targeted poisoning attack involves an adversary injecting carefully crafted malicious data or model updates into the training process to cause the model to misclassify specific inputs with target labels while maintaining the overall performance for other inputs. A stealthier attack is the backdoor (Trojan) attack that embeds a hidden trigger into the training data or model so that the model behaves normally on benign inputs but misbehaves in a controlled way whenever the trigger is present  ~\cite{10102921, 10973081, sagduyu2023vulnerabilities}. 

In this paper, we consider targeted poisoning and backdoor attacks on DFL in vehicular networks, evaluate their impact on the anomaly detection performance and develop defense mechanisms for these attacks. Our findings demonstrate that with a smart selection of features to attack, both attacks are effective, while backdoor attacks have more powerful damaging impact on the learning process, without the need for poisoning many samples. However, we show that DFL is more robust against adversaries compared to the individual learning case, which requires an adversary to attack more resources to be as effective. After highlighting the threats of targeted data poisoning and backdoor attacks, we introduce defense mechanisms based on clustering and statistical analysis that improve the probability of correctly detecting these stealthy attacks. We also evaluate the run time of the underlying computational mechanisms.

The novel contributions of this paper are multifold:
\begin{enumerate}
\item \textit{Adversary Design and Vulnerability Characterization in DFL setting}: We design targeted attacks, namely  targeted poisoning and backdoor attacks, and  perform a detailed analysis on the impact of these attacks on DFL, highlighting its vulnerabilities.
\item \textit{Performance Analysis}: Using the VeReMi Extension Dataset \cite{VeremiExtension} in comprehensive experiments, we quantify the performance degradation due to these targeted attacks and further show that DFL shows more resilience when these attacks are launched.
\item \textit{Development of Defense Mechanisms}: We present defense mechanisms against targeted attacks and showcase the improvement in inhibiting actions of adversaries when DFL is carried out in vehicular networks.
\end{enumerate}

The remainder of the paper is organized as follows. Sec.~\ref{sec:systemModel} introduces system model and the DFL architecture. Sec.~\ref{sec:Attacks} presents attacks and evaluates their impact. Sec.~\ref{sec:Defense} discusses defense mechanisms. Sec.~\ref{sec:Conclusions} concludes the paper.

\section{System Model}
\label{sec:systemModel}
We consider a multi-hop vehicular network as shown in Fig.~\ref{fig:intro}. Each vehicle sends basic safety messages (BSMs) that are periodic wireless messages containing a vehicle's real-time position, speed, and acceleration. BSMs allow other vehicles to understand traffic conditions and receive hazard alerts. These messages may include information for threat detection, alongside features like convoy location tracking and control for military communications. Each vehicle runs a deep learning algorithm that takes received BSM features from neighbor vehicles as input and classifies each message as Malicious ($M$) or Benign ($B$). As part of DFL, every participating vehicle directly shares their model weights with their neighbors, eliminating a centralized aggregator for model exchange. Each vehicle acts as an autonomous node and keeps their own data, which help preserve privacy and save communication bandwidth in congested environments. %\rev{, ultimately allowing for scaling and efficient resource usage}. 
The ultimate goal of the DFL is to detect whether a node is under Denial of Service (DoS) attack, which is revealed by $M$ or $B$ labels in the training dataset. An attacker is either external that gained access to the network or an internally disguised node, where victims could be multiple nodes within the network. DFL comprises rounds of training, where in each round every node trains their available data locally and independently. Then, each node shares its locally trained neural network model parameter with their immediate neighbors using a vehicle-to-vehicle (V2V) standard. Next, nodes aggregate model weights from their neighbors and perform federated averaging to integrate them into their own local models. Through this process, network knowledge disseminates throughout the network without the need for a central coordinator. %The details of the learning algorithm is provided in~\cite{demir2025distributedfederatedlearningvehicular}.

Accurate anomaly detection in vehicular networks relies on high-quality training data, which must be diverse and multimodal, encompassing vehicular sensor readings, communication signals, images, and text logs like BSMs collected under different operating conditions. In this paper, we use VeReMi Extension dataset for our simulations~\cite{VeremiExtension} due to its rich wireless data features and training data size. The dataset consists of BSMs that are exchanged within a simulated vehicular network. %There are 94 vehicles, where each vehicle has a different size of data  depending on their connectivity. 
There are 100 vehicles (nodes) in the vehicular network, 94 of them with (training and test) data and 6 of them without data.  The BSMs include both malfunctions (non-malicious errors from faulty sensors) and attacks (intentional disruptions). Each message is tagged with type of message, timestamp of transmission and reception, sender identity, position, speed, acceleration, and heading (with and without sensor errors). There are a total of 64,779 samples, and each sample has 22 features~\cite{ercanVeReMiDoS}. We use the morning peak traffic dataset and focus on the DoS attacks in terms of high-frequency message flooding. The connectivity of nodes is determined by the adjacency matrix that changes over time with node mobility, as studied in~\cite{demir2025distributedfederatedlearningvehicular}. %\rev{The mobility paths of the nodes resemble an organic city layout, which helps generalize our findings rather than focusing on a specific mobility pattern.} 

Each node uses an identical deep neural network architecture for attack detection that has 2 hidden layers with the sizes of 128 and 32. Each hidden layer is followed by a ReLU activation function. The output layer is a softmax function with the size of 2, indicating malicious or benign cases. Through the DFL process, nodes collectively train the model that is used for anomaly detection. We conducted our DFL, attack and defense experiments using the Keras framework with TensorFlow backend on Nvidia GeForce RTX 3060.

\section{Attacks on Distributed Federated Learning}
\label{sec:Attacks}

In this section, we assess the vulnerability of DFL in a vehicular network to targeted poisoning and backdoor attacks across various network configurations. We assume the adversary’s primary objective is to deceive the vehicular network into classifying samples labeled as \(M\) as \(B\).
%Thus, we introduce the performance metrics given in Table~\ref{tab:perforMetrics}. 
When there is no attack, $P(Y_{\mathrm{pred}}|Y_{\mathrm{true}})$ %$P(X|Y)$ 
is the probability that the DFL predicts label $Y_{\mathrm{pred}}$ given the ground truth label is $Y_{\mathrm{true}}$, e.g., $Y_{\mathrm{pred}} = B$ and $Y_{\mathrm{true}}=M$ refers to missed detection of malicious input samples. %The subscripts $trg$ and $trj$ refer to the cases when the network is attacked by targeted poisoning or backdoors, respectively. If there is no subscript, there is no attack in that case. 
When there is a targeted poisoning attack, $P_{\mathrm{trg}}(Y_{\mathrm{pred}}|Y_{\mathrm{true}})$ is the probability of predicting label $Y_{\mathrm{pred}}$ given the ground truth label is $Y_{\mathrm{true}}$. Similarly, when there is a backdoor attack, $P_{\mathrm{trj}}(Y_{\mathrm{pred}}|Y_{\mathrm{true}})$ is the probability of predicting $Y_{\mathrm{pred}}$ given the ground truth label is $Y_{\mathrm{true}}$.

\subsection{Targeted Poisoning Attacks}
\begin{figure}[!t]
\centering
  \includegraphics[width=0.55\linewidth]{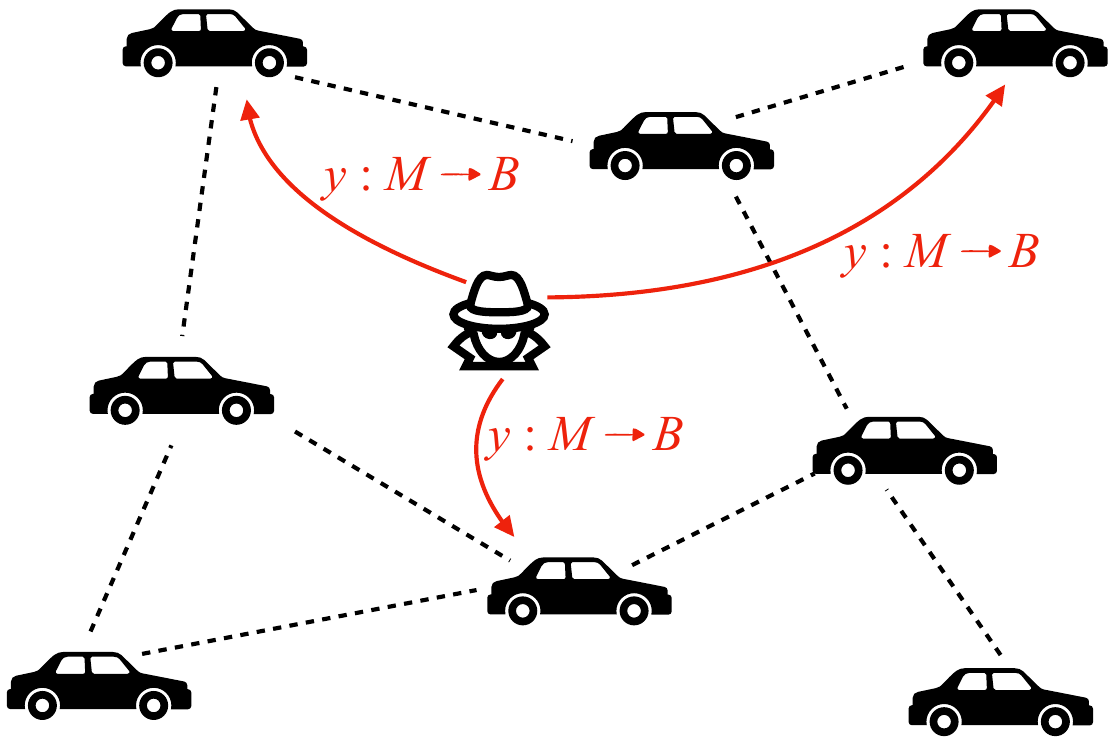}
    \caption {Targeted poisoning attack. The adversary alters the labels ($y$) of nodes from Malicious ($M$) to Benign ($B$) (i.e., $y:M \rightarrow B)$.}
  \label{fig:targetedAttack}
\end{figure}

% Challenge here would be the distributed nature of FL. When you distribute you get better results and resilience against, but it will also be more difficult to realize stealthy attacks.
In the targeted poisoning attack, an adversary deliberately flips (some of) the training labels from $M$ to $B$ to prevent the network from detecting incoming attacks. We assume the attacker has access to labels that it tries to poison. We illustrate this attack scheme in Fig.~\ref{fig:targetedAttack}, where $y$ represents training labels. The adversary conducts this attack with different attack intensities by targeting a subset of nodes in the network. For our evaluation, we define different classes of subsets that are based on network properties. Suppose $\mathcal{N}$ is the set of nodes with (training and test) data,  $\boldsymbol{d}$ is the average incoming node degree, where $\boldsymbol{d}_i$ is the average number of incoming connections to that node $i \in \mathcal{N}$ over time, $\boldsymbol{C}$ is the average size of connected components for nodes, where $\boldsymbol{C}_i$ is the average size of connected components that node $i \in \mathcal{N}$ belongs to over time, and  $\boldsymbol{c}$ is the ratio of connected times for nodes, where $\boldsymbol{c}_i$ is the ratio of times when node $i \in \mathcal{N}$ is connected to at least one other node over time.

\subsubsection{Evaluation of Targeted Attack against Individual Learning}
\label{subsubsec:singleNodeAttack}
To form the basis of evaluating the impact of targeted poisoning attacks, we start with individual learning experiments with a single node. The performance results are provided in Table~\ref{tab:TargetedSingleNode}. In this table, $p_a$ refers to the probability of attack (a label being flipped) to the targeted node. We provide probabilities of predicting labels $B$ when the actual labels are $M$ and $B$ under targeted poisoning attacks, namely $P_{\mathrm{trg}}(B|M)$ and $P_{\mathrm{trg}}(B|B)$. As the intensity of the attack increases (higher $p_a$), predictions shift towards $B$, making the attack more successful. $P_{\mathrm{trg}}(B|B)$ remains high in the presence of attack which increases the attack stealthiness.  $P_{\mathrm{trg}}(M|M)$ slightly decreases with increasing attack intensity. $P_{\mathrm{trg}}(M|B)$ remains approximately the same up to a certain intensity of the attack and begins to decrease slightly afterwards. Targeted attacks are realized instantaneously, so the only significant time portion for the run time is training and testing periods, namely, $14.77$s and $13.56$s, respectively.

\begin{table}[!t]
\footnotesize
    \centering
    \caption{Threat detection performance of individual learning under targeted poisoning attack.}
    % \vspace{-0.2cm}
    \label{tab:TargetedSingleNode}
    % First row of subtables
    \begin{tabular}{l|c|c|c|c}
        \toprule
         $p_a$ & $P_{\mathrm{trg}}(B|M)$ & $P_{\mathrm{trg}}(B|B)$ & $P_{\mathrm{trg}}(M|M)$ & $P_{\mathrm{trg}}(M|B)$ \\
        \hline
        0   & 0.2314 & 0.8308 & 0.7685 & 0.1691 \\
        0.1 & 0.2922 & 0.7928 & 0.7077 & 0.2071 \\
        0.2 & 0.3434 & 0.8033 & 0.6565 & 0.1966 \\
        0.3 & 0.3719 & 0.7928 & 0.6280 & 0.2071 \\
        0.4 & 0.4402 & 0.8985 & 0.5597 & 0.1014 \\
        0.5 & 0.5445 & 0.9450 & 0.4554 & 0.0549 \\
        0.6 & 0.5635 & 0.9238 & 0.4364 & 0.0761 \\
        0.7 & 0.5730 & 0.9598 & 0.4269 & 0.0401 \\
        0.8 & 0.5863 & 0.9767 & 0.4136 & 0.0232 \\
        0.9 & 0.6565 & 0.9978 & 0.3434 & 0.0021 \\
        1.0 & 1.0000 & 1.0000 & 0.0000 & 0.0000 \\
        \bottomrule
    \end{tabular}
\end{table}

\subsubsection{Evaluation of Targeted Attack against DFL}
The effects of targeted poisoning attack with various attack parameters are given in Table~\ref{tab:TargetedResults}, where $TopK$ refers to the set of nodes that has higher ranking in the subset designated in the subscript. The attack and its impact metrics are the same as given in Sec.~\ref{subsubsec:singleNodeAttack}. %, i.e. $p_a$, $P_{\mathrm{trg}}(B|M)$ and $P_{\mathrm{trg}}(B|B)$.
Note that, similar to the attack case against individual learning, as the attack intensity and the number of attacked nodes increase, predictions shift towards $B$, making the attack successful in all attack scenarios. The only significant time portion for the run time is training and testing periods, which are $4.62$ and $5.38$s, respectively, per training round, on average. We train the model for 201 rounds.

\subsection{Backdoor Attacks}
\label{subsec:TrojanAttack}

\begin{figure}
\centering
  \includegraphics[width=0.75\linewidth]{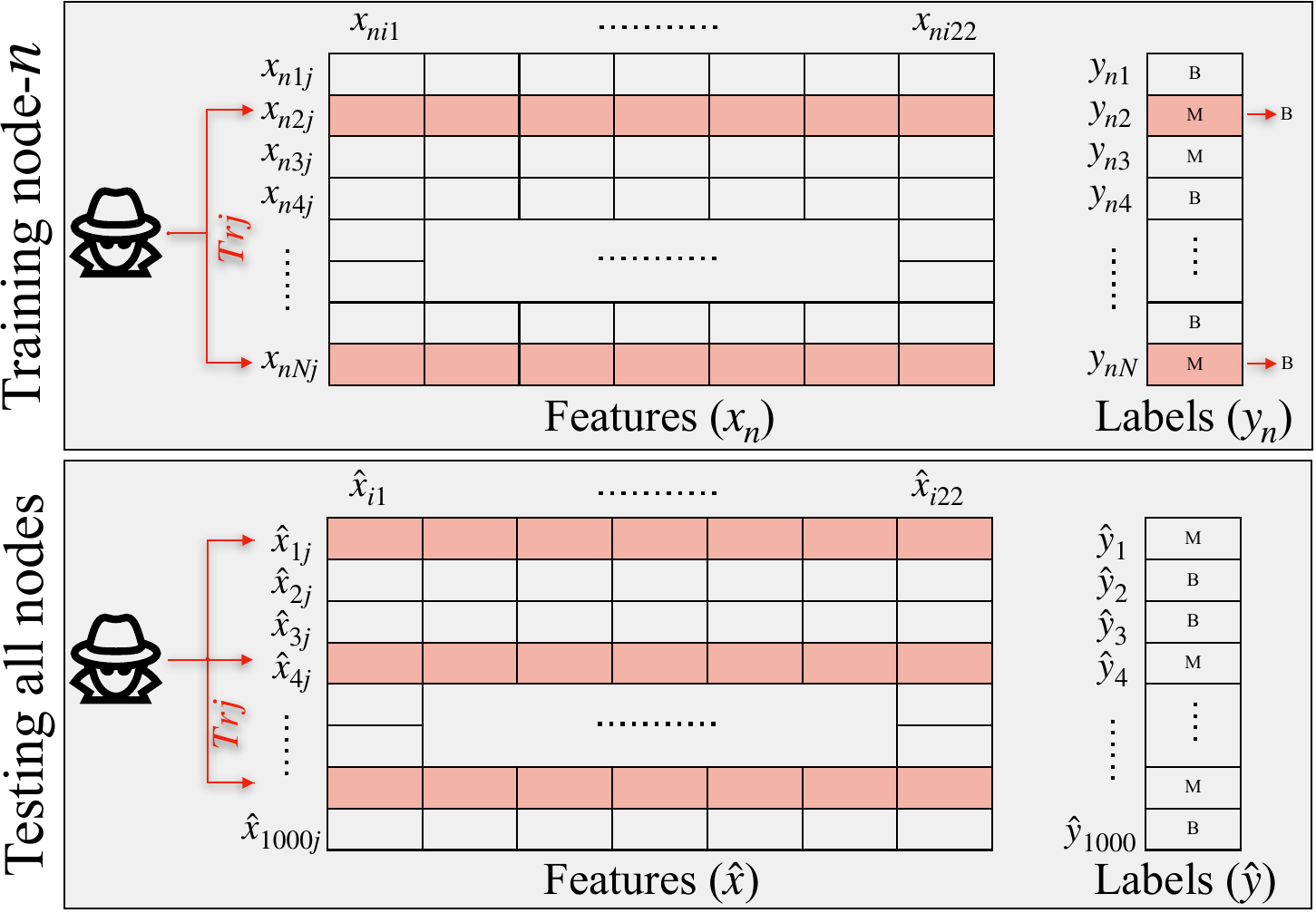}
    \caption {Backdoor attack, where the adversary selectively alters the input features $x$ and the corresponding labels $y$ of node-$n$ from $M$ to $B$ during training (upper figure). The attacked samples are colored red (e.g., $i\in\{2,N\}$). We manipulate features $(\mathrm{pos}_1,\mathrm{pos}_2)$ ($2$ out of $22$ input features, $j$) in our experiments that may target one or multiple nodes. During the test time (lower figure), triggers are added to the features ($\hat{x}$), e.g., $i\in\{1,4,999\}$, to evaluate the performance of attack that aims to result in ($\hat{y}$).}
  \label{fig:TrojanAttack}
  %\vspace{-25pt}
\end{figure}

In backdoor attacks, the adversary inserts \textit{triggers} in the training dataset by selectively altering both training data features and the corresponding labels. Later, during test time, the adversary activates these triggers that force the learned model to incorrectly classify $M$ as $B$. The attack scheme is shown in Fig.~\ref{fig:TrojanAttack}. %The adversary poisons the selected features by adding what is called \textit{triggers}.

\subsubsection{Evaluation of Backdoor Attack against Individual Learning}
\label{subsubsec:singleNodeTrojanEval}

In this subsection, we evaluate the learning performance of individual learning under backdoor attack. The single node for this evaluation is chosen as the top performing vehicle with highest accuracy in individual learning. The attack performance results are given in Table~\ref{tab:TrojanSingleNode} using different \textit{trigger selection ratios}. We use the position parameter in the input features as the trigger, which means the position features are altered as $(\mathrm{pos}_1,\mathrm{pos}_2)_{\mathrm{trj}}$, e.g., $(0,0)$, to realize the attack. $\mathrm{pos}_1$ and $\mathrm{pos}_2$ correspond to the node positions in x and y coordinates, respectively. To determine the number of triggers for a node, we calculate the distance of each training data sample $(\mathrm{pos}_1,\mathrm{pos}_2)$ to the $(\mathrm{pos}_1,\mathrm{pos}_2)_{\mathrm{trj}}$ coordinates, rank them from furthest to lowest distance and take a certain percentage of that list from the highest values, i.e. from the furthest. We increase these percentage values as shown in the \textit{Selection ratio} column in Table~\ref{tab:TrojanSingleNode}. It is evident that even attacking a small portion of the available samples of the node is enough to make the backdoor attack successful, e.g., $P_{\mathrm{trj}}(B|M)$ reaches $90\%$ even by only poisoning $<10\%$ of its training data. In the following subsection, we show how DFL provides resilience against backdoor attacks. Similar to the targeted attack case, backdoor attacks happen instantaneously. Thus, the training and testing times are the same.

\begin{table}[t]
\footnotesize
    \centering
    \caption{Effects of targeted poisoning attacks based on network properties.}
    % \vspace{-0.2cm}
    \label{tab:TargetedResults}
    % First row of subtables
    \begin{subtable}[t]{0.5\textwidth}
        \centering
        \caption{Node degree with attack parameters $TopK_d$ and $p_a$.}
        \label{tab:Targeted_sub1}
    \begin{tabular}{l|c|W{c}{13mm}|W{c}{13mm}|W{c}{13mm}|W{c}{13mm}}
        \toprule
        % \multicolumn{6}{l}{Parameters} \\
        % \hline
        $K$ & $p_a$ & $P_{\mathrm{trg}}(B|M)$ & $P_{\mathrm{trg}}(B|B)$ & $P_{\mathrm{trg}}(M|M)$ & $P_{\mathrm{trg}}(M|B)$\\
        \hline
        $25$ & $0.5$ & 0.4618 & 0.7458 & 0.5381 & 0.2541 \\
        $25$ & $1.0$ & 0.5508 & 0.7535 & 0.4491 & 0.2464 \\
        $47$ & $0.5$ & 0.5450 & 0.7757 & 0.4549 & 0.2242 \\
        $47$ & $1.0$ & 0.8077 & 0.6719 & 0.3280 & 0.1922 \\
        $70$ & $0.5$ & 0.8397 & 0.6787 & 0.3212 & 0.1602 \\
        $70$ & $1.0$ & 0.8450 & 0.9068 & 0.1549 & 0.0931 \\
        \bottomrule
    \end{tabular}
    \end{subtable}%
    \\
    \hspace{0.75cm}
    \begin{subtable}[t]{0.5\textwidth}
        \centering
        \caption{Average connected component size with attack parameters $TopK_C$ and $p_a$.}
        \label{tab:Targeted_sub2}
        \begin{tabular}{l|c|W{c}{13mm}|W{c}{13mm}|W{c}{13mm}|W{c}{13mm}}
        \toprule
        % \multicolumn{6}{l}{Parameters} \\
        % \hline
        $K$ & $p_a$ & $P_{\mathrm{trg}}(B|M)$ & $P_{\mathrm{trg}}(B|B)$ & $P_{\mathrm{trg}}(M|M)$ & $P_{\mathrm{trg}}(M|B)$\\
        \hline
        $25$ & $0.5$ & 0.4757 & 0.7475 & 0.5242 & 0.2524 \\
        $25$ & $1.0$ & 0.5614 & 0.7625 & 0.4385 & 0.2374\\
        $47$ & $0.5$ & 0.5824 & 0.7932 & 0.4175 & 0.2067\\
        $47$ & $1.0$ & 0.7039 & 0.8174 & 0.2960 & 0.1825\\
        $70$ & $0.5$ & 0.6890 & 0.8491 & 0.3109 & 0.1508\\
        $70$ & $1.0$ & 0.8365 & 0.9030 & 0.1634 & 0.0969\\
        \bottomrule
    \end{tabular}
    \end{subtable}
        \\
    \hspace{0.75cm}
    \begin{subtable}[t]{0.5\textwidth}
        \centering
        \caption{Connected time ratio with attack parameters $TopK_c$ and $p_a$.}
        \label{tab:Targeted_sub3}
        \begin{tabular}{l|c|W{c}{13mm}|W{c}{13mm}|W{c}{13mm}|W{c}{13mm}}
        \toprule
        % \multicolumn{6}{l}{Parameters} \\
        % \hline
        $K$ & $p_a$ & $P_{\mathrm{trg}}(B|M)$ & $P_{\mathrm{trg}}(B|B)$ & $P_{\mathrm{trg}}(M|M)$ & $P_{\mathrm{trg}}(M|B)$\\
        \hline
        $25$ & $0.5$ & 0.4589 & 0.7445 & 0.5410 & 0.2554  \\
        $25$ & $1.0$ & 0.5419 & 0.7550 & 0.4580 & 0.2449\\
        $47$ & $0.5$ & 0.5138 & 0.7834 & 0.4861 & 0.2165\\
        $47$ & $1.0$ & 0.6989 & 0.8296 & 0.3010 & 0.1703\\
        $70$ & $0.5$ & 0.6506 & 0.8499 & 0.3493 & 0.1500 \\
        $70$ & $1.0$ & 0.8556 & 0.9197 & 0.1443 & 0.0802\\
        \bottomrule
    \end{tabular}
    \end{subtable}
\end{table}

\begin{table}[!t]
\footnotesize
    \centering
    \caption{Threat detection performance of a individual learning under backdoor attack.}
    % \vspace{-0.2cm}
    \label{tab:TrojanSingleNode}
    % First row of subtables
    \begin{tabular}{l|c|c|c}
        \toprule
         Selection ratio & $P_{\mathrm{trj}}(B|M)$ & $P(B|M)$ & $P(M|B)$ \\
        \hline
        0     & 0.0151 & 0.2675 & 0.1501 \\
        0.005 & 0.0322 & 0.2163 & 0.2346 \\
        0.01  & 0.0815 & 0.1688 & 0.2684 \\
        0.015 & 0.1499 & 0.1631 & 0.2938 \\
        0.02  & 0.3795 & 0.2277 & 0.2621 \\
        0.0225& 0.2751 & 0.3586 & 0.0972 \\
        0.025 & 0.2637 & 0.2713 & 0.1818 \\
        0.0275& 0.5787 & 0.3055 & 0.1839 \\
        0.03  & 0.8349 & 0.3111 & 0.1670 \\
        0.04  & 0.8804 & 0.2656 & 0.2262 \\
        0.06  & 0.8519 & 0.3055 & 0.1268 \\
        0.08  & 0.8880 & 0.3187 & 0.1331 \\
        0.1   & 0.9392 & 0.2599 & 0.1585 \\
        0.2   & 1.0000 & 0.2713 & 0.2262 \\
        0.3   & 0.9905 & 0.3453 & 0.0993 \\
        0.4   & 1.0000 & 0.5313 & 0.0782 \\
        0.5   & 1.0000 & 0.6527 & 0.0401 \\
        \bottomrule
    \end{tabular}
\end{table}

%\subsubsection{Multi Node Backdoor Attack Evaluation}
\subsubsection{Evaluation of Backdoor Attack against DFL}
\label{subsubsec:multiNodeTrojanEval}

As in the individual learning case, we tested two triggers, $(\mathrm{pos}_1,\mathrm{pos}_2)_{\mathrm{trj}} = (0,0)$. The triggers are inserted during the training time and activated during the test time, as shown in Fig~\ref{fig:TrojanAttack}. During training time, we rank the input samples of nodes based on the distance between the trigger point and the received $(\mathrm{pos}_1,\mathrm{pos}_2)$.  
Then we add triggers to the samples of 10\%, 15\%, and 20\% furthest nodes, while flipping the labels from $M$ to $B$.  
During test time, the adversary activates triggers in $(\mathrm{pos}_1,\mathrm{pos}_2)$ features, where test labels are $M$.

Note that the trigger values are unknown in advance, which prevents it from being detected readily. It can only be detected after the attack continues for a while. Additionally, $(\mathrm{pos}_1,\mathrm{pos}_2)_{\mathrm{trj}} = (0,0)$ is within the simulation area. This makes the trigger more stealthy, requiring longer observations over time for defense mechanisms.

\begin{figure}
\centering
  \includegraphics[width=0.75\linewidth]{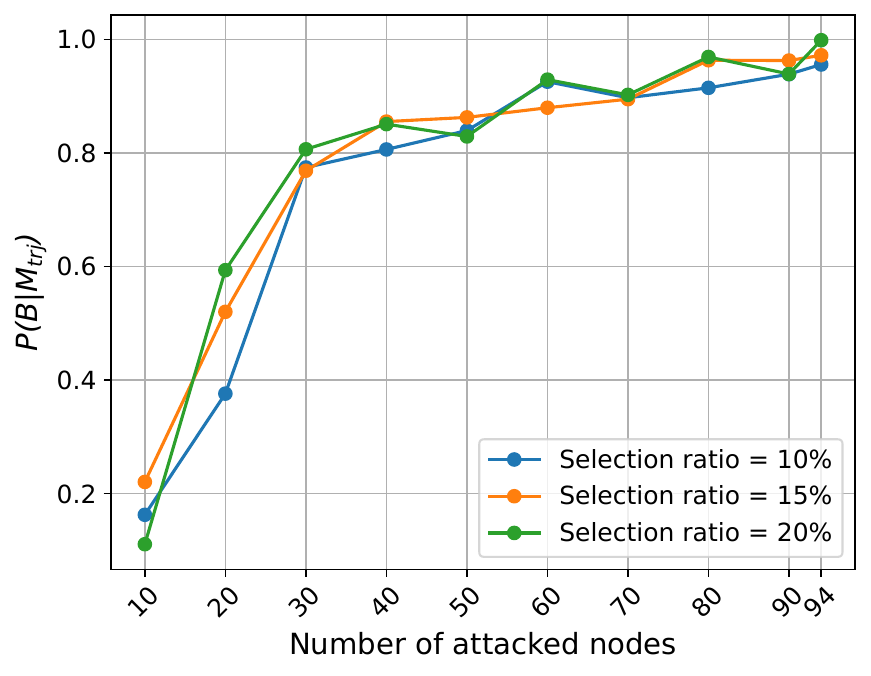}
    \caption {Finer position-based backdoor attack results.}
    %\vspace{-2mm}
  \label{fig:DetailedTrojanResults}
\end{figure}

We detail the effects of position-based backdoor attacks against DFL in Fig.~\ref{fig:DetailedTrojanResults}. The performance metric, $P_{\mathrm{trj}}(B|M)$, reaches $90\%$, averaged over all the nodes in the network, when 70 nodes are attacked with 10\%, 15\%, and 20\% of training data is added triggers. With individual learning, given in Table~\ref{tab:TrojanSingleNode}, it only takes one node to attack to reach same level of fooling. For a stark comparison, in the DFL case, even if the adversary attacks 10 nodes, the average fooling rate barely exceeds $20\%$. Thus, considering that an adversary has a limited attack budget, DFL provides substantial resilience against backdoor attacks.

\section{Defense against Adversaries in Distributed Federated Learning}
\label{sec:Defense}

\subsection{Defense against Targeted Poisoning Attacks}
\label{subsec:defense_targeted}

\begin{figure}
\centering
  \includegraphics[width=0.75\linewidth]{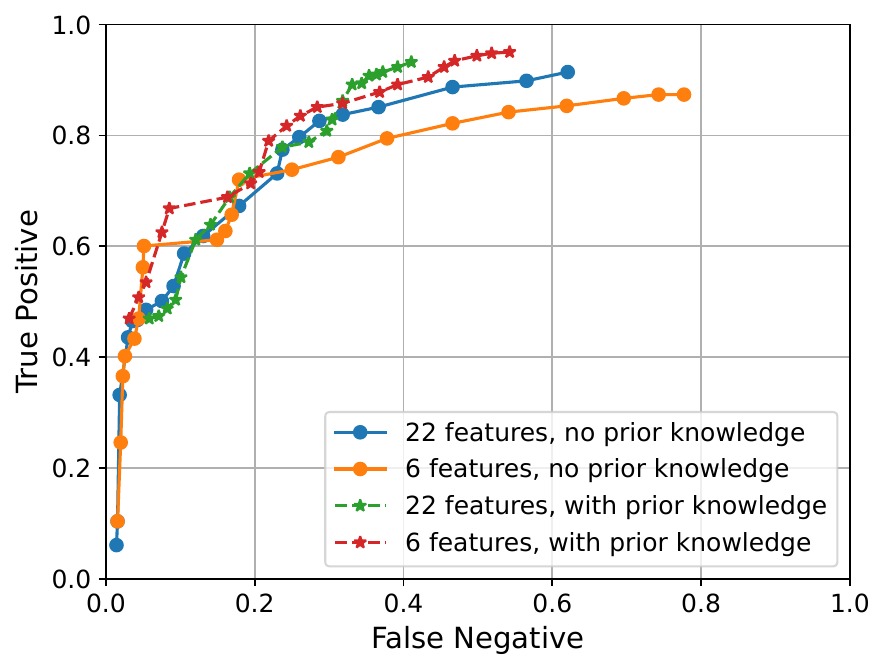}
    \caption {Targeted poisoning attack FN vs. TP curves with and without prior knowledge on statistics.}
  \label{fig:targetedDefTogether}
\end{figure}

In this subsection, we propose a clustering-based defense against targeted poisoning attacks by applying \emph{k}-means clustering ($k=2$) to each node’s samples. First, we reduce the feature set to two dimensions using principal component analysis (PCA). Next, \emph{k}-means clustering identifies two centroids, denoted \(M_{\text{center}}\) for malicious samples and \(B_{\text{center}}\) for benign samples. We classify a sample as poisoned if a benign-labeled point lies closer to \(M_{\text{center}}\) than to \(B_{\text{center}}\); otherwise, it remains classified as benign. Once poisoned samples are detected, we either remove them or correct their labels (i.e., flip them) before the next round of DFL training, choosing the action based on detection confidence. Accurately identifying benign samples is particularly critical in this setting, since attackers exploit benign labels to deceive the victim into believing that the communicating vehicle is not malicious.

In Fig.~\ref{fig:targetedDefTogether}, we present true positive (TP, correctly identifying attacked samples) and false negative (FN, incorrectly labeling a $B$ sample as attacked) curves for two scenarios. In the first scenario, the defense mechanism has prior knowledge of the sample labels before the targeted poisoning attack; in the second, it does not. We evaluate two feature‐selection schemes: (i) all 22 original features and (ii) a reduced set of six features obtained by combining the first 16 features (e.g., angle of arrival, number of received messages, etc.) \cite{ercanVeReMiDoS}. Overall, 443 of the 1582 total samples have been attacked.

Performance is evaluated for decision‐boundary distances from $M_{\mathrm{center}}$, varying between 0.5 and 1.5 times the midpoint between $M_{\mathrm{center}}$ and $B_{\mathrm{center}}$. Using $22$ and $6$ features, the defense takes $0.92s$ and $0.90s$, respectively.

Results shown in Fig.~\ref{fig:targetedDefTogether} (with and without prior knowledge on centroid) indicate that, as the decision boundary moves farther from $M_{\mathrm{center}}$, the defense captures more malicious samples (higher TP) but also incurs higher FN, illustrating a trade‐off in choosing the boundary distance. Although prior knowledge increases the maximum TP slightly, in the no‐prior‐knowledge case the minimum TP decreases and FN rises; overall, having prior knowledge reduces FN for a given TP.

\begin{figure}
\centering
  \includegraphics[width=0.65\linewidth]{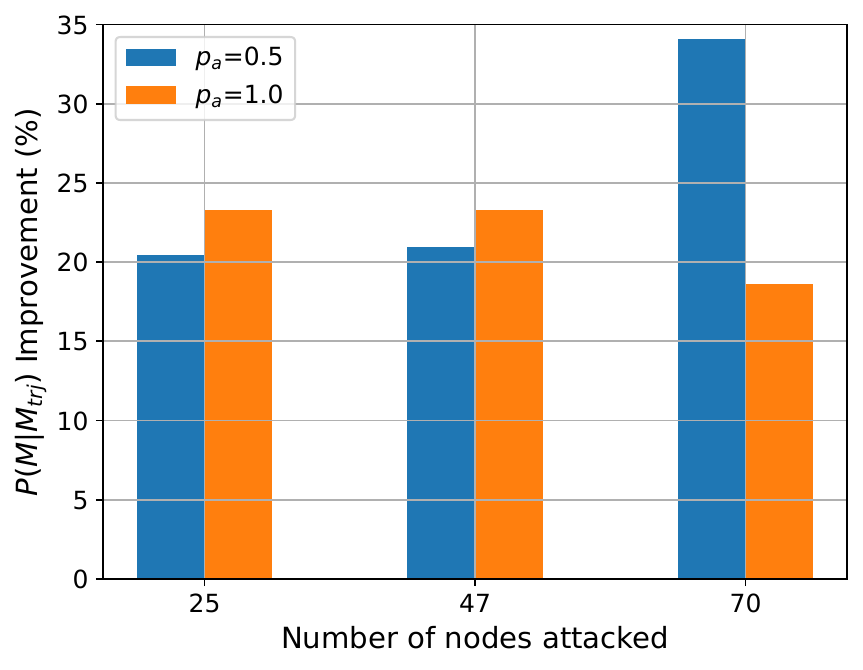}
    \caption {Improvement in detecting malicious samples under targeted attacks that was provided by the clustering defense mechanisms.}
  \label{fig:targetedDefMultiNodeTrain}
\end{figure}

To see the effect of this defense on the DFL performance, after the poisoned samples are detected using the 22-feature prior knowledge method with the decision-boundary in the middle, which gives $0.78$ and $0.27$ TP and FN values, respectively, they are removed from each node’s training dataset. Then, regular DFL procedure is followed. For the evaluation, we conducted experiments consistent with Table~\ref{tab:Targeted_sub1}, evaluating the effect of targeted attacks with 25, 47, and 70 nodes, each case having 2 different $p_a$, namely $0.5$ and $1.0$. Fig.~\ref{fig:targetedDefMultiNodeTrain} displays the improvement in $P_{trj}(M \mid M)$, which reveals that the clustering based defense is effective.

\subsection{Defense against Backdoor Attacks}
\label{subsec:defense_trojan}

\begin{figure}
\centering
  \includegraphics[width=0.7\linewidth]{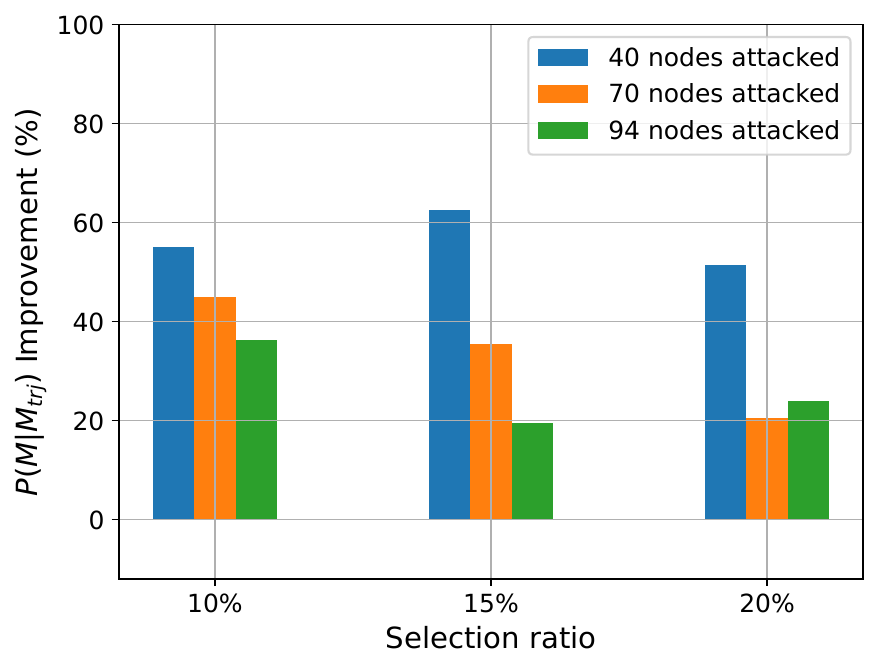}
    \caption {Improvement in detecting malicious samples under backdoor attacks that was provided by the MAD defense mechanisms.}
  \label{fig:trojanDefMAD}
\end{figure}

To defend against backdoor attacks, we adapt the median absolute deviation (MAD)–based method from \cite{davaslioglu2019trojan}. MAD quantifies the dispersion of a dataset: for all data points $x_i \in x$, $\mathrm{MAD} = \mathrm{Median}(\lvert x_i - \hat{x}\rvert)$, where $\hat{x} = \mathrm{Median}(x)$. We compute each point’s anomaly index as $\lvert x_i - \hat{x}\rvert /\mathrm{MAD}$, and, under the assumption of normally distributed data, multiply by $1.4826$ to achieve consistency with the standard deviation. A point whose normalized anomaly index exceeds $2$ has over a $95\%$ probability of being an outlier. Consequently, any label with an anomaly index above $2$ is flagged as an outlier (i.e., attacked) and removed from the training set. We then train the federated learning model on the remaining data. In our implementation, we focus on the $(\mathrm{pos}_1,\mathrm{pos}_2)$ features, which exhibit higher vulnerability compared to the other features.

In Fig.~\ref{fig:trojanDefMAD}, we illustrate the improvement in detecting malicious labels, i.e.\ $P_{\mathrm{trj}}(M|M)$, achieved by the MAD defense method. To evaluate this strategy, we adopt the network configuration from Table~\ref{tab:Targeted_sub1}, in which nodes are attacked according to their node‐degree ranking since the average fooling probability, $P_{\mathrm{trg}}(B|M)$, is higher in this scenario than in other targeted‐attack cases. The MAD‐based defense yields a $38.8\%$ improvement over the no‐defense baseline, averaged across all attack intensities. MAD takes only $0.53s$ to estimate attacked samples in the entire dataset.

\section{Conclusion}
\label{sec:Conclusions}
We examined the benefits, advantages, vulnerabilities, and resilience of DFL when applied to vehicular networks, which have critical tactical and commercial applications. Unlike traditional, server‐dependent FL, DFL enables participating vehicles to train models locally and share updates directly, conserving bandwidth in congested environments and reducing power consumption in remote battlefields. Using the VeReMi Extension dataset, we demonstrate that DFL significantly enhances the resilience of a vehicular network compared to individual (local) learning under sophisticated adversarial machine learning attacks. In particular, DFL is influential in reducing the impact of both targeted poisoning and stealth backdoor attacks in a network tasked with detecting cyberattacks. To further bolster DFL’s resilience, we introduce clustering‐ and statistical‐analysis–based defense mechanisms. The clustering approach against targeted poisoning and the MAD‐based statistical method against backdoor attacks improves malicious‐label detection accuracy by an average of $23\%$ and $38\%$, respectively, across the network. Future work will focus on refining these defense mechanisms in real‐device implementations.

\section*{Acknowledgements}

Research was sponsored by the Army Research Laboratory under RTX BBN Technologies, Inc. subcontract and was accomplished under Cooperative Agreement Number W911NF-24-2-0131. The views and conclusions contained in this document are those of the authors and should not be interpreted as representing the official policies, either expressed or implied, of the Army Research Laboratory or the U.S. Government. The U.S. Government is authorized to reproduce and distribute reprints for Government purposes notwithstanding any copyright notation herein.

\bibliographystyle{IEEEtran}
\bibliography{References}

\end{document}